\documentclass[twocolumn,prl,showpacs,aps,superscriptaddress]{revtex4-2}

\usepackage{amsmath}
\usepackage{multirow}
\usepackage{graphicx} % for figures
\usepackage{color}
\usepackage[latin1]{inputenc}
\usepackage{enumitem}
\usepackage[french]{babel}
\usepackage{ulem}
\begin{document}

\date{\today}

\title{Condensation mechanism of high-$T_c$ cuprates\,: the key role of pairon excitations}

\author{Yves Noat$^*$}

\affiliation{Institut des Nanosciences de Paris, CNRS, UMR 7588 \\
Sorbonne Universit\'{e}, Facult\'{e} des Sciences et Ing\'{e}nierie, 4 place
Jussieu, 75005 Paris, France}

\author{Alain Mauger}

\affiliation{Institut de Min\'{e}ralogie, de Physique des Mat\'{e}riaux et
de Cosmochimie, CNRS, UMR 7590, \\ Sorbonne Universit\'{e}, Facult\'{e} des
Sciences et Ing\'{e}nierie, 4 place Jussieu, 75005 Paris, France}

\author{William Sacks}

\affiliation{Institut de Min\'{e}ralogie, de Physique des Mat\'{e}riaux et
de Cosmochimie, CNRS, UMR 7590, \\ Sorbonne Universit\'{e}, Facult\'{e} des
Sciences et Ing\'{e}nierie, 4 place Jussieu, 75005 Paris, France}

\affiliation{Research Institute for Interdisciplinary Science,
Okayama University, Okayama 700-8530, Japan}

%\pacs{74.72.h,74.20.Mn,74.20.Fg}


%\large
\begin{abstract}
In this article we show that the condensation mechanism in cuprates
involves the strong coupling of the condensate to pairon excited
states. We present an accessible formalism that significantly
extends our previous work, providing a theoretical basis for the
energy-dependent gap function $\Delta(E)$. The latter is
proportional to the effective spin exchange energy, $J_{eff}$, with
no retardation effects, such as the case of spin-fluctuation or
phonon mediated couplings. The fundamental parameters of the
superconducting (SC) state are the condensation energy per pair,
$\beta_c$, and the antinodal energy gap, $\Delta_p$, which are
quantitatively extracted by fitting the cuprate quasiparticle
spectrum from tunneling experiments.
\\ An explicit formula for the critical temperature is
also derived in the model. Valid for any doping, we find $T_c$ to be
proportional to $\beta_c$, and not the gap $\Delta_p$, in sharp
contrast to conventional SC. The numerical factor
$\beta_c/k_BT_c\simeq 2.24$ originates from pair excitations of the
condensate, following Bose statistics, with a mini-gap $\delta_M
\simeq 1\,$meV in the excitation spectrum.
These results strongly suggest that the same `all-electron' mechanism    is at work all along the
$T_c$-dome.\\
$^*$\,Corresponding author: yves.noat@insp.jussieu.fr
\end{abstract}

\maketitle

\subsection{Introduction}

Given the wide variety of theoretical models of high-$T_c$
superconductivity \cite{PhysicaC_Singh2021}, spanning many decades,
prompts the more general question: {\it what constitutes a
satisfactory or acceptable model of a given physical phenomenon?}

{This question has been seriously considered across scientific
disciplines, suggesting three general properties of a working theory:
First, a successful theory must be in agreement with a wide range of
independent non-contradictory experimental facts -- a basic tenet of
the scientific method. Secondly, as proposed by William of Occam
\cite{Occam}: the simplest theory is to be preferred -- a profound
idea that has often emerged throughout history. For example,
Einstein's theory of special relativity is based on a single axiom:
that the speed of light {\it in vacuo} is a constant independent of
the observer, and the consequences follow. Thirdly, following the
philosophy of Karl Popper \cite{Popper1959}, a successful theory must be
refutable \cite{Note_Popper}. On the contrary, an erroneous model or theory that is
difficult to refute lingers, obscuring the truth. Finally,
successful theories are often predictive, adding confidence in their
validity.}

{Examining the case of conventional superconductivity provides a
relevant and very good example to address these criteria. Moreover,
pointing out the key highlights is very helpful to better understand our
approach to the high-$T_c$ problem.}

More than four decades after the discovery of superconductivity by
Kamerlingh Onnes, the ground breaking paper of Bardeen, Cooper,
Schrieffer (BCS) successfully explained this fundamental quantum
phenomenon \cite{PR_BCS1957}. Prior to BCS, other phenomenological
models had been proposed, but the BCS theory rapidly gained wide
acceptance, and a plethora of experiments followed to validate it
\cite{Parks}, satisfying our first criterion. An essential
ingredient of their microscopic theory, illustrating our second
criterion, is the existence of Cooper pairs: bound states of
electrons near the Fermi level \cite{PR_Cooper1956}, expressed as
singlet states in $k$-space: $|\vec{k} \uparrow -\vec{k} \downarrow
\rangle$. Below the critical temperature, the metallic system
condenses into a coherent quantum state of Cooper pairs.

The original paper of BCS\,\cite{PR_BCS1957} gives the essential and
necessary properties of a superconducting material, i.e. zero
resistivity, perfect diamagnetism, and the discontinuity in the
specific heat. The theory predicts a temperature-dependent energy
gap $\Delta(T)$, the order parameter of the transition; its
zero-temperature value being directly proportional to the transition
temperature: $\Delta(0)=1.76\,k_B\,T_c$. Moreover, the theory
establishes the precise shape of the quasiparticle (QP) density of
states (DOS) $N_S(E)$, measured at low temperature in a
superconductor-insulator-normal metal tunnel junction, or between a
metallic tip and a superconducting material. The QP spectrum was
first directly measured by I. Giaever in 1962 \cite{PR_Giaever1962},
confirming the BCS energy gap at the Fermi level.

At finite temperatures, quasiparticle (fermionic) excitations grow,
resulting in a decrease of the energy gap $\Delta(T)$, which
finally vanishes at $T_c$. The interpretation is thermal
pair-breaking, and the empirical shape of $\Delta(T)$ closely
follows the prediction of the BCS gap equation \cite{PR_Douglass}.
The presence of the gap is also confirmed by the specific heat,
which increases exponentially with temperature, and exhibits a
characteristic discontinuity at $T_c$ where the gap vanishes, in
excellent agreement with the BCS
theory\,\cite{PR_Brown1953,PR_Corak1956}.

The discovery of signatures beyond the gap in the tunneling spectra,
associated with the strong coupling of electrons with phonons
\cite{PR_Giaever1962}, further strengthened the theory. Prior to BCS, the isotope effect, wherein $T_c$ was known to vary as $T_c \propto M^{-1/2}$ where $M$ is its atomic mass, pointed to the key role of phonons \cite{PR_Maxell1950,PR_Reynolds1950}. However, the physical
mechanism of the electron-phonon coupling was decisively confirmed
by Rowell and McMillan who showed that the measured quasiparticle
DOS could be compared with the calculated one deduced from the real
phonon spectrum \cite{PRL_McMillan1965} -- a triumph of condensed
matter physics.

From these considerations, our three criteria are well satisfied in
the case of conventional superconductivity. The multitude of
quantitative agreements between the BCS theoretical predictions and
experimental results supports its validity beyond a reasonable doubt.
% Figure 1
\begin{figure}
\includegraphics[trim=10 10 10 10, clip, width=8.0 cm]{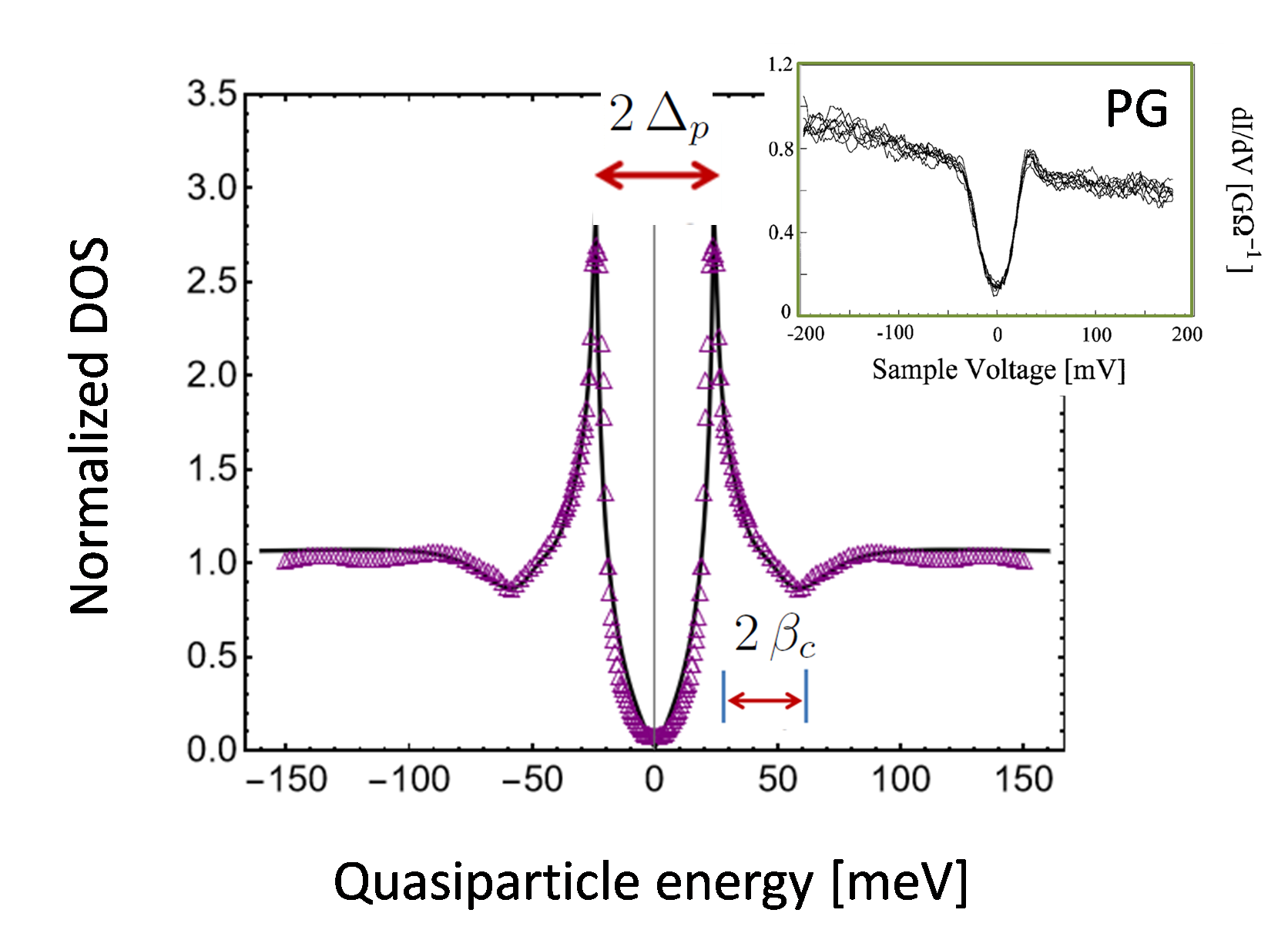}
\caption{(Color online) Quasiparticle spectrum (purple triangles)
measured by tunneling spectroscopy in
Bi$_2$Sr$_2$CaCu$_2$O$_{8+\delta}$ (data taken from Ref.\,\cite{PRL_Fang2006}; the background has been substracted and the spectrum is symetrized, corresponding to the unoccupied states' side) and corresponding fit
using the gap equation Eq.\,(\ref{Eq_gapeq4}) (black line). A clear
`dip' is visible at the energy $\sim 2\beta_c$ above the gap energy
$\Delta_p$. Inset: Quasiparticle spectrum on a disordered film in a
{\it non-superconducting} region\,: the quasiparticle peaks are
largely attenuated and the dips are absent, indicating the loss of
superconducting coherence (adapted from \cite{PRL_Cren2000}).}
\label{Fig_Spectrum}
\end{figure}

What is the situation in the case of high-$T_c$ cuprates? Nearly
four decades after their discovery by  Bednorz and M\"{u}ller
\cite{ZPhys_Bednorz1986}, there is still no consensus on a single
mechanism which coherently explains the pair formation, the
$T_c$-dome, and the complex phase diagram as a function of hole or
electron doping. In the context of the pairing
mechanism and the unconventional phase diagram, many valuable
theoretical approaches have been proposed, including
Refs.\,\cite{Sci_Anderson1987,Nat_Emery,PRB_varma,PRL_Huscroft1998,
PRL_Curty2002, JPhysConf_Newns2007, JChemPhys_Tahir-Kheli2010,
PRL_Gull2013, RevMod_Chen2024, PhysicaB_Marino2025}. To this day,
with respect to the criteria we discussed at the outset, the
validity of their conclusions remains to be confirmed.

The quasiparticle excitation spectrum is obviously unconventional (see
Fig.\,\ref{Fig_Spectrum}).  At low energy (compared to the SC peak
gap value), it reveals the $d$-wave nature of the condensate
wavefunction. At higher energy, a dip is seen above the SC peak, a
phenomenon that has attracted much attention
(\cite{Revmod_Fisher2007} and refs. therein). The precise shape of the wide quasiparticle peaks, followed by the sharp dips, in the tunneling experiments, is
challenging to describe theoretically.
A popular model is the strong coupling to a collective mode,
which was used to describe, with some success, the quasiparticle DOS
 \cite{PRL_Ahmadi2011, PRB_Berthod2013}.
However, the difficulty to fit properly the data, over a wide doping range, and the number of parameters needed, suggest
that the issue remains an open question.

Contrary to conventional superconductivity, a gap persists in the vortex core \cite{PRL_Pan2000}, in disordered films
where superconducting coherence is lost \cite{PRL_Cren2000}
(Fig.\,\ref{Fig_Spectrum}, inset), or at the critical temperature
\cite{PRL_renner1998_T,JphysSocJap_Sekine2016}. The gap energy
decreases linearly as a function of hole concentration and does not
follow the $T_c$-dome
\cite{JPhysSocJap_Nakano1998,RepProgPhys_Hufner2008,PRB_Sacks2006}.
Hence, the energy gap cannot be the order parameter, a fact that
refutes the BCS theory applied to cuprates.

A commonly held idea is that the mechanism changes as a function of
carrier concentration, throughout the SC dome, from unconventional
superconductivity in the underdoped regime towards conventional BCS
superconductivity in the overdoped regime. In this framework, the proposed mechanism is due to a
Bose-Einstein condensation in the underdoped regime (small coherence
length) while a BCS-like condensation is recovered in the overdoped
regime (large coherence length). However, this scenario has not been
proven and, for example, the measurements of the SC coherence length
by Wang et al. (\cite{EPL_Wen2003},\cite{EPL_Wang2008}) suggest
otherwise.

As noted by Uemura \cite{PhysicaC_Uemura1997}, in this `BEC-BCS'
crossover picture, the interaction responsible for superconductivity
should evolve from a non-retarded one at low doping towards a
retarded one, the BCS case, on the overdoped side. However, this is
empirically not the case: the tunneling spectra have essentially the
same shape all along the $T_c$-dome, strongly indicating that the
same mechanism is at work \cite{PRL_Miyakawa1999}. Precise fits of
the quasiparticle spectrum throughout the $T_c$-dome demonstrate
unambiguously that the interaction is instantaneous
\cite{PRB_Sacks2006, SciTech_Sacks2015}, confirming the findings of recent works based on a microscopic models where the key role of double occupancy constraints \cite{PRB_Zinni2021} and the effect of spin fluctuations were explored \cite{NewJPhys_Yamase2023}.

{Clearly, the case of high-$T_c$ cuprates remains elusive. In what
follows, we present a straightforward theory of the cuprate
mechanism which matches the empirical facts simply and accurately.
Based on the initial idea of `pairons', i.e. bound hole pairs due to
their local antiferromagnetic environment, the comprehensive model
described here satisfies all three of the essential criteria
discussed above. In particular, our theory has the third essential
property, according to Popper, of being refutable.}

\subsection{The pairon model: a mechanism for cuprates}

In the spirit of Occam's razor, the complex phase diagram of
cuprates can be understood with few assumptions. In our model, inspired by early numerical works in the 1990s (see \cite{RevMod_Dagatto1994} and references therein), hole pairs on adjacent copper sites, or `pairons', form on the characteristic scale equal to the
AF correlation length \cite{EPL_Sacks2017}. Pairons are one example of preformed pairs. They are described in real space while Cooper pairs are described in $k$ space. However, there is a direct connection between the two, leading to the electronic gap at the Fermi level. The pairon binding
energy $\Delta_p$, which is also the antinodal gap energy, is given
by the effective antiferromagnetic exchange energy $J_{eff}$. The
antinodal gap $\Delta_p$ decreases linearly as a function of the
number of holes \cite{EPL_Sacks2017}, and the associated pseudogap
temperature $T^*$ extrapolates to the N\'eel temperature at zero
doping \cite{PRB_Cyr-Choiniere2018}.

% Figure 2
\begin{figure}[b]
\includegraphics[trim=10 10 10 10, clip, width=6.5 cm]{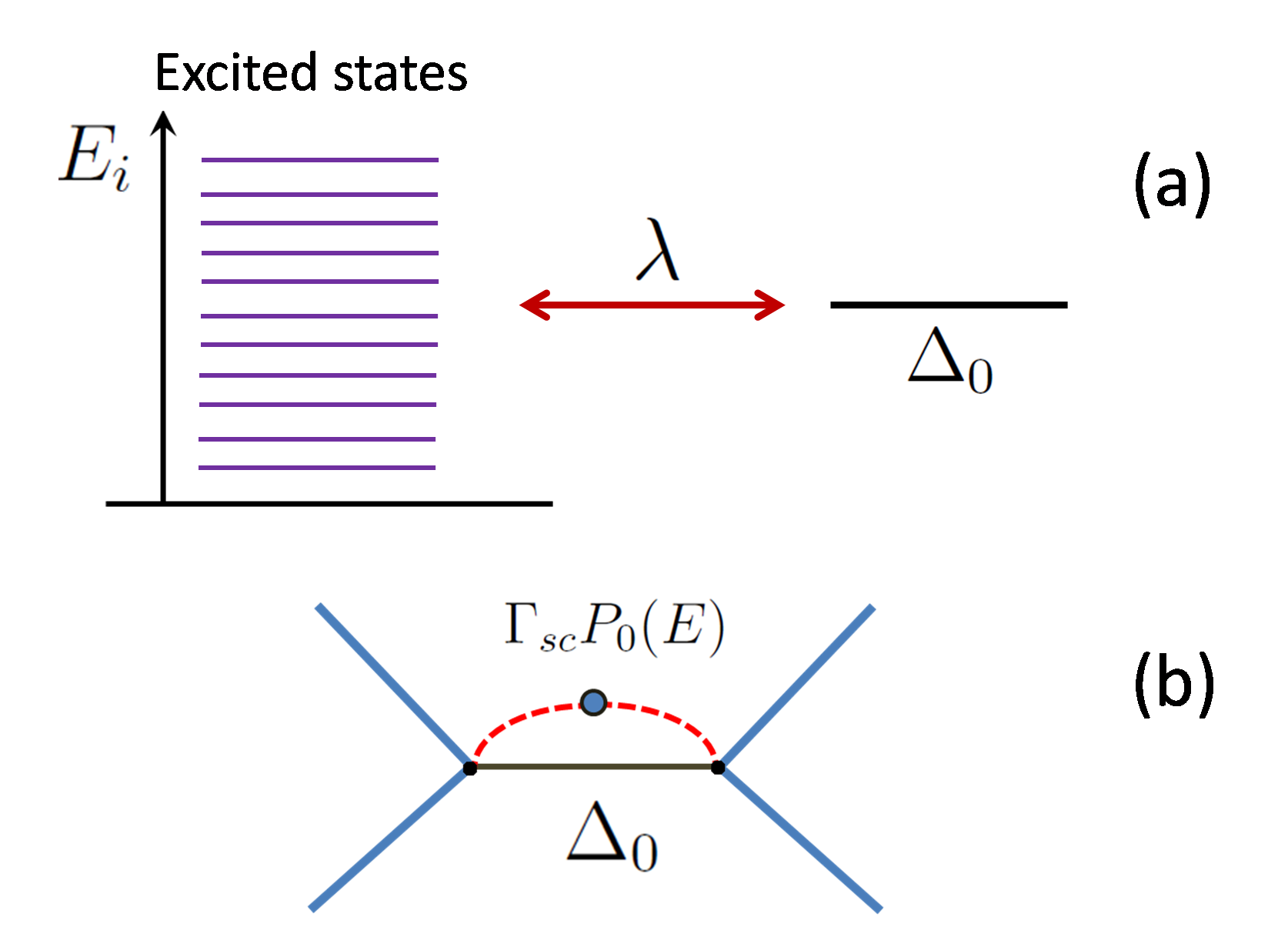}
\caption{(Color online) (a) Coupling of the non-interacting ground
state with energy $\Delta_0$ to a continuum of excited pair states
$E_i$, with a coupling constant $\lambda$. (b) Illustration of the
resulting self-consistent gap function, Eq.\,(\ref{Eq_gapeq3}).}
\label{Fig_Coupling}
\end{figure}

Below $T_c$, pairons condense into a macroscopic quantum state.  A
second characteristic energy, $\beta_c$, must be considered, which
is the condensation energy (per pair). While $\Delta_p$ follows
$T^*$, the pairon formation temperature, $\beta_c$ is proportional
to the critical temperature and originates from the interaction
between pairons. The dependence of these two parameters on the
carrier concentration ($p$) is linked to the geometrical constraints
imposed on holes (and pairons) on the copper sites of the CuO plane
\cite{PhysLettA_Noat2022,ModelSimul_Noat2023}. In particular, it
proves that $\beta_c$ follows a dome in the phase diagram (we omit its $p$ dependence, unless needed).

The pairon model successfully describes the measurements of
tunneling spectroscopy \cite{EPJB_Sacks2016,Jphys_Sacks2017}, ARPES
\cite{Jphys_Sacks2018,EPL_Noat2019}, magnetic susceptibility
\cite{SolStatCom_Noat2021}, specific heat
\cite{SolStatCom_Noat2021,SolstatCom_Noat2024}, and the upper
critical field \cite{PhysLettA_Noat2025}. {In these works, the
pairon excitations, in addition to quasiparticles, play a central
role. The reason for this is that $T_c$ (equivalently $\beta_c$) is
always below $T^*$ (equivalently $\Delta_p$), in the temperature
(energy) phase diagram. In other words, when SC coherence is lost,
the condensate transitions to the pseudogap state of incoherent
pairons (see inset in Fig.\,\ref{Fig_Spectrum}).}

{In this work, a straightforward quantum formalism is proposed to
calculate the self-consistent coupling with pair excitations to
finally express this unconventional condensate. The methodology
proposed here is an extension of our previous work, providing
additional rigor and understanding of the role of the pair excited
states (\cite{Jphys_Sacks2017, EPL_Sacks2017}, and refs therein).
The condensate fundamental parameters are shown to be contained in
the energy-dependent gap function $\Delta(E)$. The latter allows to precisely fit the spectra measured by
low-temperature tunneling spectroscopy \cite{PRB_Sacks2006, Jphys_Sacks2017, SciTech_Sacks2015}, from which the important
parameters, the energy gap $\Delta_p$ and the condensation energy
$\beta_c$ are deduced.}

The model also accounts for thermal excitations and provides an
explicit and compact formula for the critical temperature. From the
experiments, in agreement with previous work, we show that $\beta_c$
is proportional to $T_c$ for a wide doping range. Moreover, we show
that the numerical factor $\beta_c/k_B\,T_c\simeq 2.24$ is related
to the pair excitations, following Bose statistics, and to the
presence of a mini-gap $\delta_M$ in the excitation spectrum.
Finally, we conclude that all three energy parameters are
proportional to the static spin exchange energy, $J_{eff}$.

\subsection{Condensation mechanism}

{As mentioned previously, inherent in the pairon model is the
existence of pair excited states, a degree of freedom which is
unconventional and absent in the BCS theory. This section is devoted
to describing how this coupling arises.}

\vskip 2mm

\centerline{\it Formalism for excited state coupling}

\vskip 2mm

We start from the non-interacting ground state of energy $\Delta_0$
coupled to a continuum of excited pair states with energies $E_i$,
as illustrated in Fig.\ref{Fig_Coupling}\,(a). The hamiltonian is $H
= H_0 + H_{ex} + H_c$ with the terms\,:
\begin{itemize}
    \item[{$\bullet$}] $H_0 = \Delta_0\, |p\rangle \langle p|$ \quad (starting condensate)
    \item[{$\bullet$}] $H_{ex} = \sum_i E_i\, |ex_i\rangle \langle ex_i|$ \quad (excited states)
    \item[{$\bullet$}] $H_c = \lambda\, \sum_i \left( |p\rangle \langle ex_i| + |ex_i\rangle \langle p| \right)$ \quad (coupling)
\end{itemize}
The eigenstates are conveniently taken as: $|p\rangle$ the
condensate, $|ex_i\rangle$ the excited pair states, while $\lambda$
is the energy coupling parameter between the condensate and the
excited states. The latter is taken independent of $i$ for
simplicity. Evidently, the states $|ex_i\rangle$ represent a degree
of freedom which is absent in the BCS theory where the elementary
excitations are only quasiparticles.

The starting Green's operator for the non-interacting ground state
is\,:
\begin{equation}
    G_0(E_i) = \frac{|p\rangle \langle p|}{E_i - \Delta_0 + i\eta}
\end{equation}
where $\eta$ is an infinitesimal, and for the excited states is\, \cite{Economou}:
\begin{equation}
    G_{ex}(E_i) = -\pi \, i \, N_{ex}(E_i) |ex_i\rangle \langle
    ex_i|  \label{Eq_exdos}
\end{equation}
where $N_{ex}(E_i)$ is the density of excited states.

To evaluate the properties
of the coupled system, we write the total Green's operator $G$ using
Dyson's equation\,:
\begin{equation}
    G = (G_0 + G_{ex}) + (G_0 + G_{ex}) H_c G
\end{equation}
The objective is to focus on the modifications of the initial
condensate, and thus to calculate the projection of Dyson's equation
onto the state $| p \rangle$. Then, defining the necessary matrix
elements,
\begin{align}
    G_{p,p} &= \langle p |\, G\, | p \rangle \\
    G_{ex_i,p} &= \langle ex_i |\, G\, | p \rangle
\end{align}
we obtain,
\begin{align}
    G_{p,p} &= g_0 + \lambda\, g_0\, G_{ex_i,p} \label{Eq_Gpp} \ \\
    G_{ex_i,p} &= \lambda\, g_{ex_i}\, G_{p,p}
\end{align}
where:
\begin{align}
   g_{ex_i} &= \langle ex_i |\, G_{ex}\, | ex_i \rangle \\
    g_0 &= \langle p |\, G_0\, | p \rangle \label{Eq_Gpp4}
\end{align}
Combining Eqs. (\ref{Eq_Gpp}) to (\ref{Eq_Gpp4}), after some
algebra, results in\,:
\begin{equation}\label{Eq_Gpp0}
    G_{p,p}(E_i) = \frac{1}{E_i - \Delta_0 - \lambda^2\, g_{ex_i}(E_i)}
\end{equation}
where the infinitesimal factor is suppressed. This `exact' Green's
function for the coupled system allows to find both the excited
state distribution and the SC final ground state.

Replacing $g_{ex_i}(E_i)$ by its expression using
Eq.\,(\ref{Eq_exdos}), we get an effective gap function having the
compact form:
\begin{equation}\label{Eq_gapeq1}
    \Delta(E_i) = \Delta_0 -\pi \, i \lambda^2\,\,
    N_{ex}(E_i)
\end{equation}

\vskip 3mm

\centerline{\it Excited pairon density of states}

\vskip 2mm

An essential feature of the model is that pairons can be excited out of the condensate: the
non-superconducting state is characterized by an incoherent
distribution of pairs, the pseudogap state. As mentioned
previously, it is manifested by the pseudogap as seen in the vortex
core or due to disorder, at low temperatures, and also just above
$T_c$ (inset in Fig.\,\ref{Fig_Spectrum}).

This effect is immediately demonstrated in our formalism by choosing
the lowest order form in the gap equation (\ref{Eq_gapeq1}): $\pi
\lambda^2\, N_{ex} = \sigma_0$. Using Eq.\,\ref{Eq_Gpp0} leads to:
\begin{equation}
    N_p(E_i) = -\frac{1}{\pi}\, \text{Im}\, G_{p,p}(E_i) =
    \frac{1}{\pi} \frac{\sigma_0}{(E_i - \Delta_0)^2 + \sigma_0^2}
\end{equation}
which is a Lorentzian distribution of width $2\sigma_0$ centered on
$E_i = \Delta_0$. This provides a simple derivation of the pairon
excited state distribution at zero-temperature, the Cooper Pair
Glass (CPG), or equivalently 'pairon glass', introduced in Ref.\,\cite{EPL_Sacks2017}.

The interpretation is quite straightforward: an excited pair is in a
non-stationary state, with a lifetime $\sim \hbar/\sigma_0$. This
corresponds to a complex pole in the Green's function at $E_{pole} =
\Delta_0 - i \sigma_0$, as seen in Fig.\,\ref{Fig_Gap_PG}, left
panel.

{The difference between the `initial' (lowest order) and the `final'
pair density of states is significant: in the latter case the peak energy parameter $\Delta_0$ is preserved. This peak energy
value coincides with the total energy ($\Delta_0 = \Delta_p +
\beta_c)$, so that the CPG state can be interpreted as the result of
an adiabatic transformation from the SC state to the incoherent
pseudogap state. If all the pairs become incoherent, this suggests
that $2\sigma_0 \sim \Delta_0$, a relation that is indeed found in
the experiments (see the phase diagram in Ref.\,\cite{EPL_Sacks2017}).
\vskip 2mm

\centerline{\it Gap function in the superconducting state}

\vskip 2mm

{In the SC ground state, the gap function (\ref{Eq_gapeq1}) has to
be a real function, i.e. the pole of the Green's function
(\ref{Eq_Gpp0}) must have no imaginary part.} This property is
strictly necessary in order to have a coherent stationary state. The
condensation must therefore be associated with a shift of the pole
in the complex plane towards the real axis (Fig.\ref{Fig_Gap_PG},
left panel). This conclusion might seem odd at first, since
(\ref{Eq_gapeq1}) appears to be a complex function.

To restore a stationary state (SC) from the non-stationary CPG
state, we introduce an imaginary coupling parameter in the Green's
function\,:
\begin{equation}
    \lambda^2 = -i\,|\lambda|^2
\end{equation}
In other words, the SC coupling involves a phase shift of $\pi/4$ in
the coupling argument. This assumption leads to the real gap
equation\,:
\begin{equation}
    \Delta(E_i) = \Delta_0 - \pi \,|\lambda|^2\,
    N_{ex}(E_i)\label{Eq_gapeq2}
\end{equation}
Introducing the initial or lowest-order excited states DOS, i.e.
$\pi \,|\lambda|^2\,N_{ex} \to \sigma_0$, would simply lead to a
constant shift of the ground state energy, $\Delta_0 \to \Delta_0 -
\sigma_0$. This lowest order DOS is
insufficient to capture the right physics, and we must look to higher order.

Instead, in the spirit of the Breit-Wigner perturbation theory, we
replace $N_{ex}(E_i)$ by the `final' DOS calculated in the previous
subsection. The meaning is clear\,: the excited state DOS in Eq.\,(\ref{Eq_gapeq2}) results from the {\it final coupling} of $|p\rangle$
to excited states $|ex_i\rangle$. The following replacement in
Eq.\,(\ref{Eq_gapeq2}) is now needed\,:
$$
    N_{ex}(E_i) \to  N_p(E_i)
$$
leading to the expression\,:
\begin{equation}
    \Delta_{SC}(E_i) = \Delta_0 - \pi \,|\lambda|^2\, N_p(E_i)
\end{equation}
where $N_p(E_i)= -(1/\pi)\,\text{Im}\,G_{p,p}(E_i)$ is the final
density of excited states discussed previously.

The physical consequence is that the condensate is now linked to the CPG
density of states\,:
\begin{equation}
    \Delta_{SC}(E_i) = \Delta_0 - |\lambda|^2\,\frac{\sigma_0}{(E_i - \Delta_0)^2 + \sigma_0^2}
\end{equation}
and the SC gap function becomes energy dependent, with a strong
resonance at the value $E_{res} = \Delta_0$.

To simplify the notation, and to better connect with our previous
work (for example \cite{EPL_Sacks2017}), we now define \(\Gamma_{SC}
\) with \( |\lambda|^2 = \Gamma_{SC}\cdot \sigma_0 \), so that:
\begin{equation}\label{Eq_gapeq3}
    \Delta_{SC}(E_i) = \Delta_0 - \Gamma_{SC}\,P_0(E_i)
\end{equation}
with the convenient dimensionless form for the distribution\,:
\begin{equation}
    P_0(E_i) = \frac{\sigma_0^2}{(E_i - \Delta_0)^2 + \sigma_0^2}
\end{equation}
{Evident in the gap equation (\ref{Eq_gapeq3}) above, at the heart
of the pairon condensation mechanism, is the role of the coupling
term related to pair excitations, proportional to $\Gamma_{SC}$, see
Fig.\,\ref{Fig_Coupling}\,(b).}

{Before proceeding, we note some important points about the energy
scale $E_i$ in this final SC gap equation. First, we evaluate the
ground state energy in the antinodal direction. This corresponds to
$E_i = \Delta_p$, where $\Delta_p$ is the antinodal gap (assuming
pure $d$-wave pairing). Using this value in the gap equation (\ref{Eq_gapeq3}), we
have the essential property that $\Delta_{SC}(\Delta_p)=\Delta_p$,
by definition of the Fermi level gap. Self-consistency requires
that\,:
\begin{equation}
    \Delta_p = \Delta_0 - \Gamma_{SC}\,P_0(\Delta_p)
\end{equation}
The second term represents the pair correlation necessary to have a
SC condensate: the coherence energy $\beta_c$. This leads to the
fundamental identity\,:
\begin{equation}
    \beta_c = \Gamma_{SC}\,P_0(\Delta_p)
\end{equation}
which determines the value of $\Gamma_{SC}$ in the ground state\,:
\begin{equation}
\Gamma_{SC} = \beta_c\,\left(1+\frac{\beta_c^2}{\sigma_0^2} \right)
\end{equation}
For reference, the ratio $\Gamma_{SC}/\beta_c$ varies from $\sim\,$
(1.0 - 2.0) in the full doping range. Finally we note that the
correlation term has the effect of lowering the Fermi level gap,
i.e. $\Delta_p < \Delta_0$. This effect can be directly seen in the
DOS as illustrated in Fig.\ref{Fig_Effect_Coupling}, right panel}.

{For energies larger than the gap, i.e. $E_i>\Delta_p$ there are two
important cases\,: at very low temperature the excited states are
{\it virtual pairs}, strongly coupled to the quasiparticles, and are
central to the tunneling DOS. On the contrary, at finite
temperature, {\it real} pairon excitations of energy $\varepsilon_i
= E_i-\Delta_p$ are the essential degrees of freedom of the system
responsable for SC phase decoherence leading to $T_c$. Both these
important cases, absent in the BCS mechanism, are described below.}

\vskip 2mm

\centerline{\it Connection to the quasiparticle DOS}

\vskip 2mm

Using the self-consistent gap function, we can calculate the final
quasiparticle DOS. Derivations of the quasiparticle DOS can be found
in our previous work, \cite{PRB_Sacks2006}, \cite{SciTech_Sacks2015}, \cite{Jphys_Sacks2017},
\cite{EPJB_Sacks2016}.

While Ref.\,\cite{Jphys_Sacks2017} employed a lengthy second-order
perturbation theory to calculate the virtual pair -- quasiparticle
coupling, a major simplification consists in imposing simple energy
conservation\,: each $E_i$ energy level above $\Delta_p$ is equal to
a QP energy, $E_i = E_k$, where $E_k$ is the quasiparticle energy in the antinodal direction, so that\,:
\begin{equation}\label{Eq_gapeq4}
    \Delta_{SC}(E_k) = \Delta_0 - \Gamma_{SC}\,P_0(E_k)
\end{equation}
with \,$E_k = \sqrt{\epsilon_k^2 +\Delta_{SC}(E_k)^2}$ and $\epsilon_k$ is the kinetic energy relative to $E_F$.

\begin{figure}
\centering
  \hspace{-5 mm}
\includegraphics[trim=10 10 10 10, clip, width=9.2 cm]{ 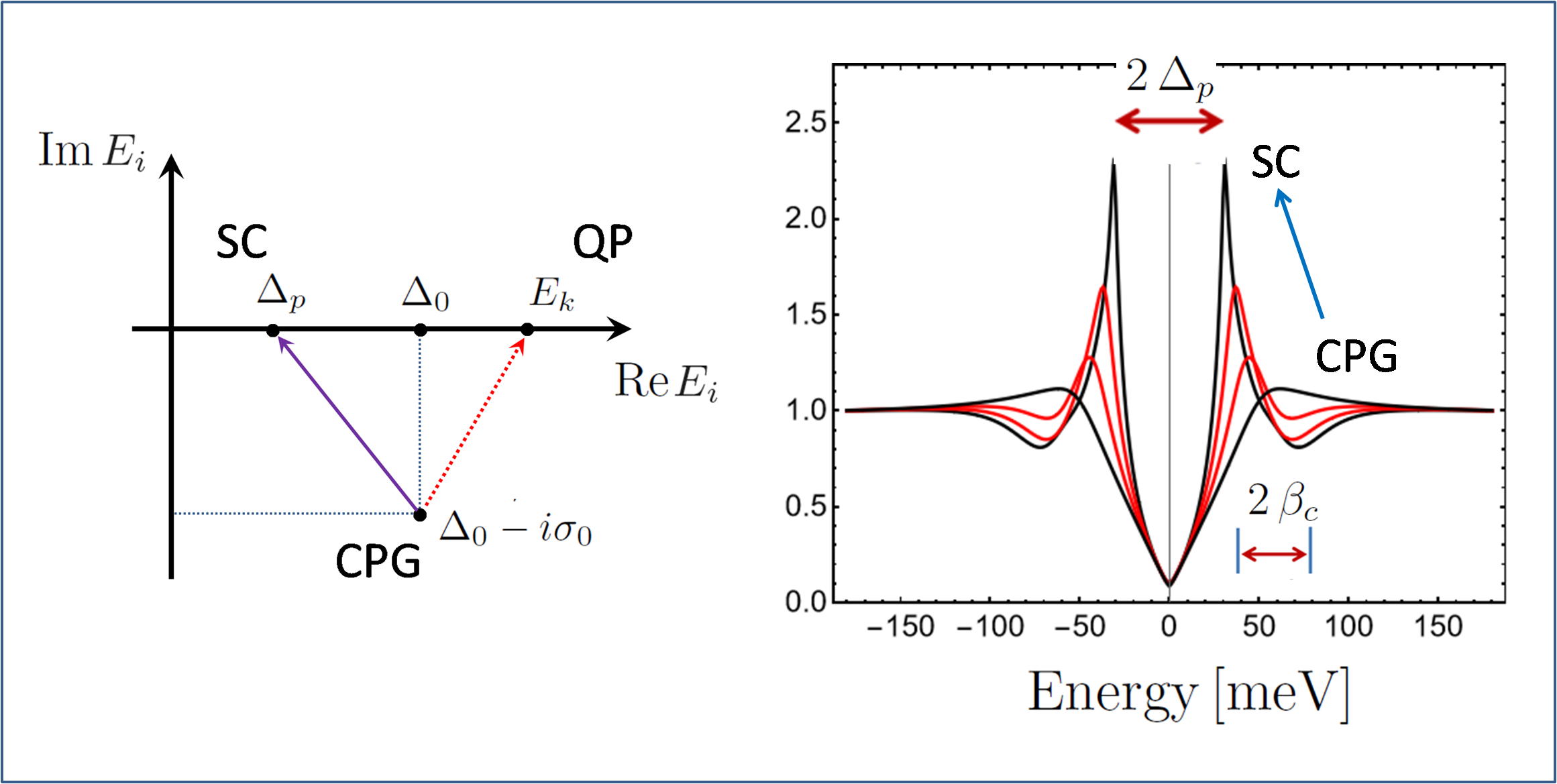}
\caption{(Color online) Left panel: Evolution of the Green's
function pole in the complex plane from the superconducting state to
the pseudogap state. Right panel: Evolution of the normalized
quasiparticle spectrum from the superconducting state to the
pseudogap states using Eq.\,(\ref{Eq_gapeqT}). In the SC state a
clear dip is visible beyond the gap energy, while in the CPG state,
the dip, as well as the coherence peaks, are absent. The latter is close
to the quasiparticle spectrum observed in vortex cores
\cite{Revmod_Fisher2007,PhysicaC_Hoogenboom2000}. }
\label{Fig_Gap_PG}
\end{figure}

\begin{figure}
\centering
  \hspace{-5 mm}
\includegraphics[trim=10 10 10 10, clip, width=9.2 cm]{ 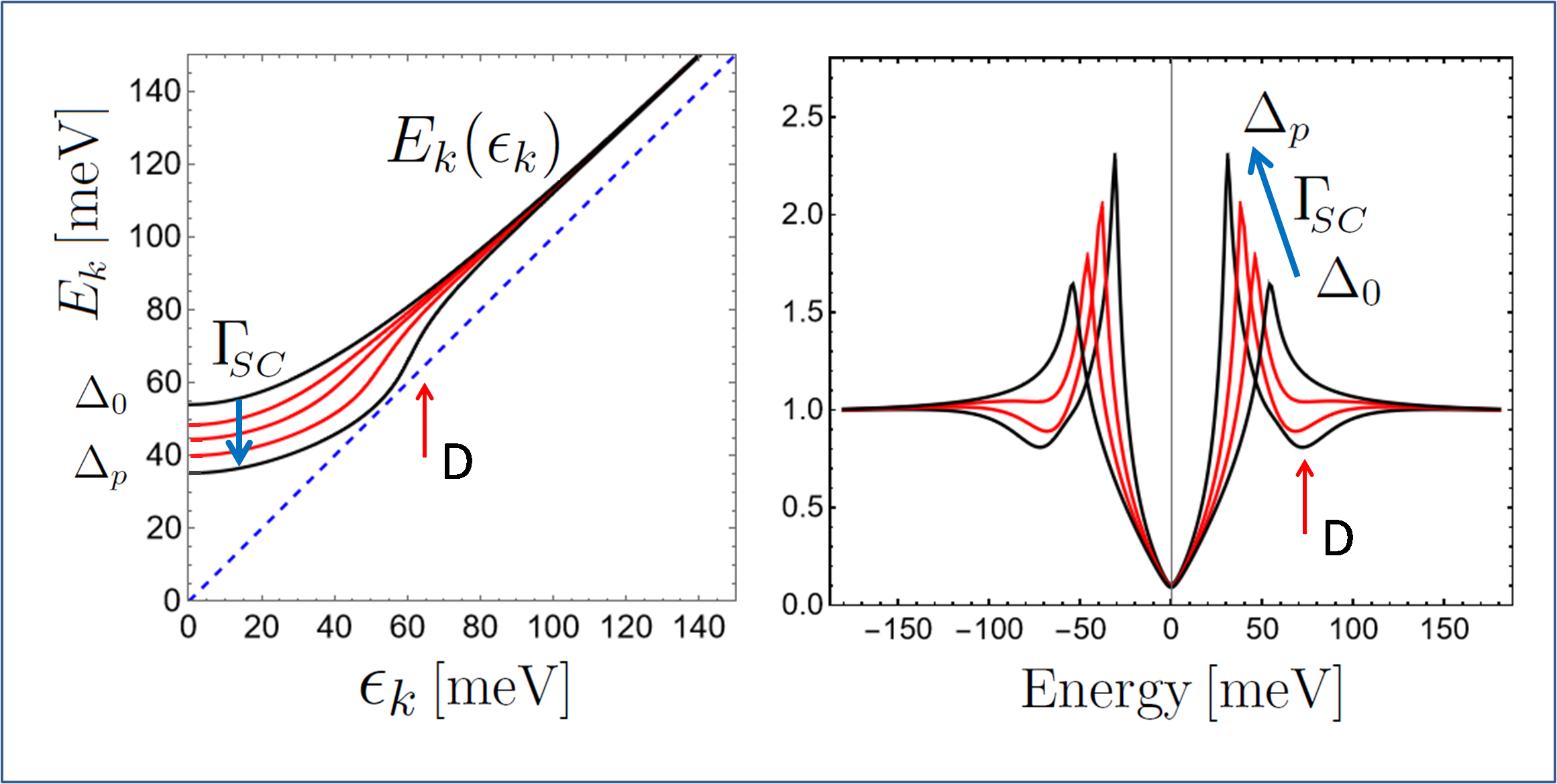}
\caption{(Color online) Left panel: Evolution of the dispersion
relation (in the antinodal direction) for increasing coupling constant $\Gamma_{SC}$ to the
excited states. One sees the smooth transition from a hyperbolic BCS
shape to the final form showing a progressively sharper kink. Right
panel: Effect of the coupling to excited states on the normalized
quasiparticle spectrum showing the appearance of the dip (indicated
by the letter `D') for increasing $\Gamma_{SC}$. Note that the
conservation of states is valid when the DOS shape evolves. In both
panels Eq.\,(\ref{Eq_gapeq4}) is used.} \label{Fig_Effect_Coupling}
\end{figure}

The gap itself now depends on the QP energy $E_k$, as first proposed
by Cren et al. \cite{EPL_Cren2000}. Finally, we are able to give a
precise meaning to this unconventional energy dependence: the
condensate-excited pair coupling. Moreover, in the above derivation,
there is no need for retardation effects resulting from
boson-mediated coupling, such as phonons or magnons.
Our findings are aligned with other more microscopic models such as in Ref.\,\cite{PRB_Zinni2021,NewJPhys_Yamase2023}.

The gap function (\ref{Eq_gapeq4}) allows to fit very precisely the
quasiparticle DOS measured by tunneling spectroscopy, solid line in
Fig.\,\ref{Fig_Spectrum}, using the methods of Ref.\,\cite{PRB_Sacks2006,SciTech_Sacks2015}. In
Fig.\,\ref{Fig_Effect_Coupling} the quasiparticle spectrum is
calculated as a function of the coupling parameter $\Gamma_{SC}$,
showing the appearance of the unconventional shape and associated
dispersion relation. As a function of the coupling parameter
$\Gamma_{SC}$ one can follow the evolution of the QP-DOS, from a pure
$d$-wave form with a large gap $\Delta_0$, to a sharper and wider form
with a peak value of $\Delta_p$. The clear dips beyond the QP peaks
are evident and due to the resonant nature of the coupling term in
the gap equation. Finally, we emphasize the unconventional final
dispersion relation (see Fig.\,\ref{Fig_Effect_Coupling}, left
panel) showing a sharp kink just above the gap energy. This effect
was also predicted in the calculated QP self-energy in
\cite{PRB_Sacks2006}.

% Figure 5
\begin{figure}[t]
\includegraphics[trim=10 10 10 10, clip, width=6.5 cm]{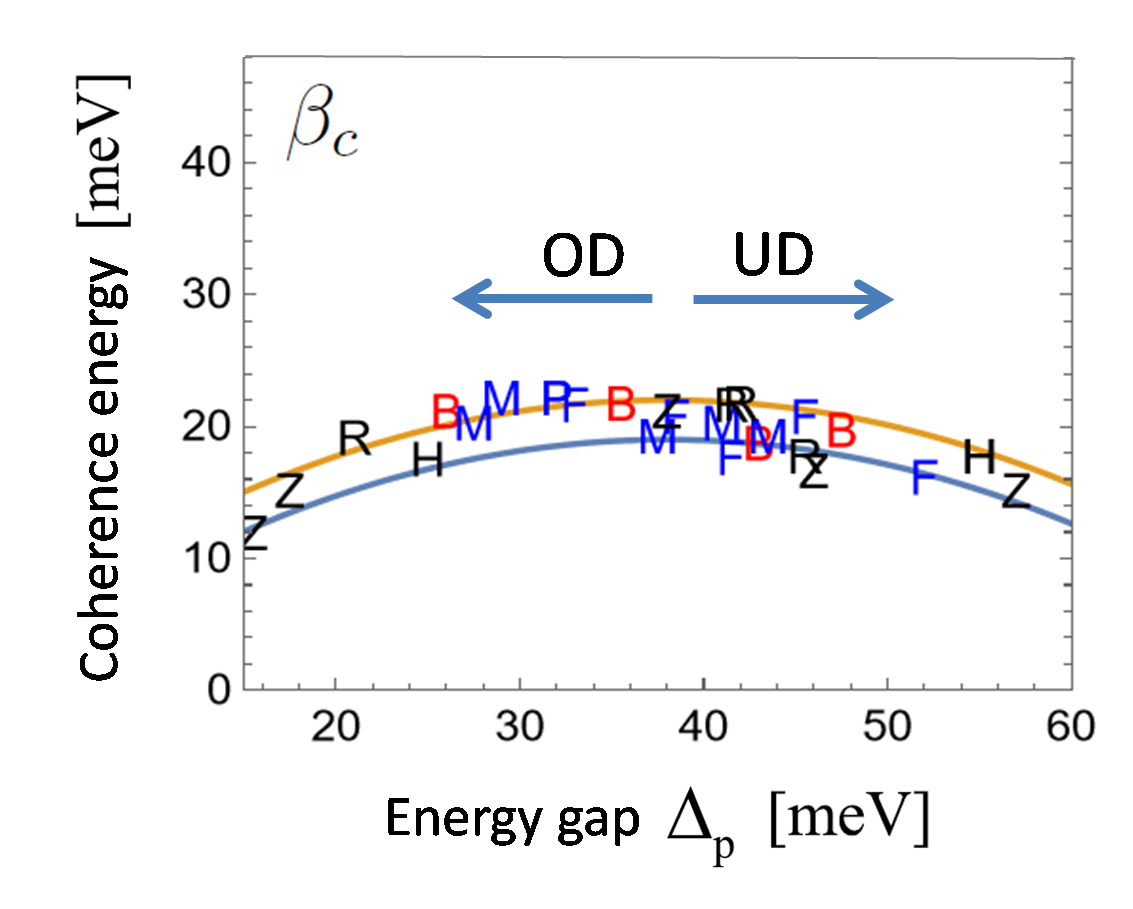}
\caption{(Color online) Condensation energy $\beta_c$ extracted from
experimental tunneling spectra on BSCCO plotted as a function of $\Delta_p$.
Solid parabolic lines are a guide to the eye using
Eq.\,(\ref{Eq_betadelta}). The points are deduced by precisely
fitting the spectra from the references indicated, using the gap
equation Eq.\,(\ref{Eq_gapeq4}). R: Renner et al.
\cite{PRL_Renner1998,Revmod_Fisher2007}, M: McElroy et al.
\cite{Sci_McElroy2005}, F: Fang et al. \cite{PRL_Fang2006}, B, C:
Cren et al. \cite{EPL_Cren2000,PRL_Cren2000}, Z: Zasadzinski et al.
\cite{PRL_Zasadzinski2001}, H: Yang He et al. \cite{Science_YHe} }
\label{Fig_Beta}
\end{figure}

It is important to note that the characteristic dip, and the sharp
peaks in the QP-DOS, are the signatures of SC coherence. These
signatures have been measured repeatedly and accurately, by
tunneling spectroscopy \cite{Revmod_Fisher2007} from which the
fundamental parameters of cuprates can be inferred
\cite{PRB_Sacks2006,SciTech_Sacks2015,Jphys_Sacks2017}. Precise fits
to the experimental spectra yield the peak-to-peak gap (2$\Delta_p$)
as well as the dip position $\sim \Delta_p + 2\beta_c$. The low-temperature values of these two parameters are summarized in
Fig.\,\ref{Fig_Beta}. Quite clearly, the coherence energy parameter
is following the SC dome. The value of $\beta_c$ at optimal doping is $\sim
J_{eff}/4 \simeq 18.5$\,meV to within $\sim \pm 5$\% for
Bi$_2$Sr$_2$CaCu$_2$O$_{8+\delta}$.

\begin{figure}[b!]
\includegraphics[trim=10 10 10 10, clip, width=4.5 cm]{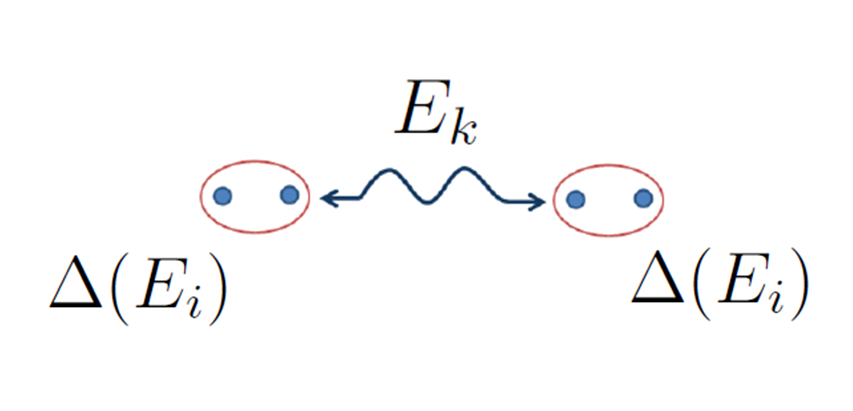}
\caption{(Color online) All fermion proposal for the quantum
coherence of the SC condensate: pairon-quasiparticle coupling. }
\label{Fig_pairing}
\end{figure}

{The gap function (\ref{Eq_gapeq4}) associated with the SC
condensate points to a more fundamental concept. Since the
self-consistent gap depends on $E_k$, and the quasiparticle
dispersion in turn depends on $\Delta(E_k)$, as in
Fig.\ref{Fig_Effect_Coupling} left panel, there must be an intimate
link between quasiparticle coupling and the pairing mechanism. This
was suggested in our previous work, \cite{Jphys_Sacks2017}. As
indicated in Fig.\ref{Fig_pairing}, we propose that the virtual
exchange of quasiparticles between pairons plays a key role leading
to the SC coherence.}

Finally, a useful expression for the SC \( \to \) CPG {\it gradual}
transition can be obtained using the total gap function:
\begin{equation}\label{Eq_gapeqT}
    \Delta_T(E_k) = \Delta_0 - \Gamma_{SC} P_0(E_k) - i\,\Gamma_{CPG}\,\sigma_0
\end{equation}
where  \( \Gamma_{CPG}\neq 0 , \Gamma_{SC} \to 0 \), in the CPG
state and \( \Gamma_{SC}\neq 0, \Gamma_{CPG} \to 0 \), in the SC
state. This complex gap function describes the zero-temperature
evolution from SC to CPG states (see Fig.\,\ref{Fig_Gap_PG}, right
panel). For example, it corresponds to the local tunneling
spectroscopy along a line on the surface which crosses a vortex
\cite{PhysicaC_Hoogenboom2000}. In the vortex core center, where SC
coherence is broken, the QP-DOS exhibits a gap without coherence
peaks and no characteristic dip.

\subsection{Critical temperature}

An important result of the BCS theory is the famous ratio of the gap
to the critical temperature. The numerical factor
$\Delta/k_BT_c=1.76$ is the mathematical consequence of the fact
that elementary excitations of the condensate are of fermionic type,
the quasiparticles. A similar relation, but resulting from a
different physical origin, can be found in cuprates between the
condensation energy and the critical temperature using our model, in
very good agreement with experiments.

\vskip 2mm

\centerline{\it Characteristic energies of the model}

\vskip 2mm

The characteristic energies of the model as a function of hole
concentration $p$ are expressed as (for
Bi$_2$Sr$_2$CaCu$_2$O$_{8+\delta}$):
\begin{align}
    \Delta_p(p) &= J_{eff}\, (1 - p') \label{Eq_Deltap}\\
    \beta_c(p) &= J_{eff}\, p'(1 - p') \label{Eq_Betac}\\
    \Delta_0(p) &= \Delta_p(p) + \beta_c(p) \\
    \sigma_0(p) &\simeq \Delta_0(p)/2
\end{align}
where $J_{eff}$ is the effective exchange energy at the dome onset
($p=p_{min}$), and $ \sigma_0$ is the width of the excited pair
energy distribution $P_0(E_i)$ \cite{EPL_Sacks2017,
ModelSimul_Noat2023, Jphys_Sacks2017}. For convenience, we use the
reduced hole concentration
$$p'=(p-p_{min})/(p_{max}-p_{min})$$ with $p_{min}=.05$ and
$p_{max}=.27$. The factors $(1 - p')$ and $p'(1 - p')$ are simply
1/2 and 1/4, respectively, at optimal doping.

The characteristics of these parameters as a
function of carrier concentration ($p$) are linked to the
geometrical constraints imposed on holes (and pairons) on the copper
sites of the CuO plane
\cite{PhysLettA_Noat2022,ModelSimul_Noat2023}. In fact, the phase
diagram reflects the statistical properties of particles on a square
lattice with a finite interaction (or correlation) distance
\cite{ModelSimul_Noat2023}. The strongest constraint is that two
pairons cannot be associated with the same site. So, for example, if
$p'$ is proportional to the pairon density, then $1-p'$ is
proportional to the number of unoccupied sites, which gives the
$p$-dependence of $\Delta_p$, Eq.\,(\ref{Eq_Deltap}). On the other hand,
$\beta_c$ depends on the statistical pairon-pairon correlation, and
is proportional to the product\,: $\beta_c \propto p'\cdot (1-p')$,
as in Eq.\,(\ref{Eq_Betac}).

In the case of Bi$_2$Sr$_2$CaCu$_2$O$_{8+\delta}$
the characteristic ratio $\beta_c/\Delta_p$ takes the simple form\,:
$$
    \frac{\beta_c}{\Delta_p} = p'
$$
which was first noted by Sacks et al.\,\cite{PRB_Sacks2006}, albeit
with a different notation ($\Delta_{\varphi}$ in lieu of $\beta_c$).
Removing $p'$ using Eq.\,(\ref{Eq_Deltap}) leads to a novel expression for
$\beta_c$\,:
\begin{equation}\label{Eq_betadelta}
    \beta_c = \Delta_p\,(1-\frac{\Delta_p}{J_{eff}})
\end{equation}
which is precisely the `dome-shape' function depicted in
Fig.\,\ref{Fig_Beta}.  At optimal doping, the values are
$\Delta_p^{opt} = J_{eff}/2$ and $\beta_c^{opt} =
J_{eff}/4$ with $J_{eff}\simeq 74$\,meV.

The case of La$_{2-x}$Sr$_x$CuO$_4$ is slightly
different due to its smaller $T_c$ and smaller $T_c/T^*$ ratio
compared to Bi$_2$Sr$_2$CaCu$_2$O$_{8+\delta}$. This is consistent
with the fact that La$_{2-x}$Sr$_x$CuO$_4$ is single layer with a
weaker interlayer coupling. The following adjustments are deduced
from experiment \cite{PRB_Cyr-Choiniere2018}\,:

\begin{align}
    \beta_c(p) &= J_{eff} p'(1 - p') \\
    \Delta_p(p) &= \alpha \, J_{eff} (1 - p')
\end{align}
where $\alpha$, a dimensionless coefficient close to unity, accounts for the effect of interlayer magnetic coupling, as mentioned above. For LSCO, we find $J_{eff} \simeq 31$\,meV and $\alpha\simeq 1.25$ (and for BSCCO, $\alpha=1$).
Here we neglect the `kink' in the $T_c$-dome on the underdoped
side, possibly due to charge ordering. The total energy equation\,:
$\Delta_0 = \Delta_p + \beta_c$ remains valid.

We can thus propose that, in the general case of
cuprates, the coherence parameter is of the general form\,:
$$
\beta_c(p) = C \cdot p'\cdot \Delta_p(p)
$$
where $C$ is of order unity, which again stresses the departure of
the pairon model from conventional BCS. Based on their study of
the low-temperature QP-DOS, Sacks et al.\,\cite{PRB_Sacks2006}
proposed this general formula for the coherence energy. A similar relation (but with $p$ instead of $p_{min}$) was previously
suggested by Ido et al. \cite{LowtempPhys_Ido1999} and Dipasupil et al. \cite{JPhysSocJap_Dipasupil2002}. This relation now finds a more solid theoretical foundation in this work.

Finally, a mini-gap, $\delta_M$, in the energy excitation spectrum is a necessary
component of the pairon model. We attribute the existence of a mini-gap, absent in a pure Bose-Einstein condensation of non-interacting bosons, to the fact that pairons are interacting on a finite distance. $\delta_M$ is the energy required to remove one pairon from the condensate to give the first excited state. This parameter is proportional to $\beta_c$, which can be understood from the
statistical approach that we proposed in \cite{PhysLettA_Noat2022}:
\begin{equation}
\delta_M(p') \approx p_{min}\ \beta_c(p')
\end{equation}
{$\delta_M(p')$ being proportional to $p_{min}$. The mini-gap is directly related to the existence of a finite pairon-pairon interaction length $d_0$, since $p_{min}=(a_0/d_0)^2$, where $a_0$ is the Cu-Cu lattice distance in the CuO plane. In practice
we have used the following expression for best results\,:
$\delta_M(p') = 1.1\times p_{min}\ J_{eff}\ p'\,(1-p') $. The main
point is that the mini-gap is proportional to $\beta_c$, and hence
follows the $T_c$ dome.

\vskip 2mm

\centerline{\it Temperature dependence of the condensation energy}

\vskip 2mm

In this section, we focus on {\it real} pairon
excitations, of energy $\varepsilon_i = E_i - \Delta_p$, as discused
in the previous section. Using the SC gap function,
Eq.\,(\ref{Eq_gapeq3}), we see that the fluctuation probability
is\,:
$$\frac{\delta \Delta_{sc}}{\beta_c} = \frac{\Gamma_{SC}\,
\overline{P_0}(\varepsilon_i)}{\beta_c}$$ Since there is a change of
variable, we write the density of pair excited states as:
\begin{equation}
    \overline{P_0}(\varepsilon_i)= \frac{\sigma_0^2}{(\varepsilon_i - \beta_c)^2 + \sigma_0^2}
\end{equation}
Including the occupation probability, $f_B(\varepsilon_i, T)$, the
number of thermally excited pairons from the condensate is given by:
\begin{equation}
    \frac{\Gamma_{SC}}{\beta_c}\, \int_{\delta_M}^{\Delta_0}\, \overline{P_0}(\varepsilon_i)\,
    f_B(\varepsilon_i, T)\, d\varepsilon_i
\end{equation}
where
$f_B(\varepsilon_i,T)=1/\left(\exp\left(\frac{\varepsilon_i-\mu_b}{k_BT}\right)-1
\right)$ is the Bose-Einstein distribution, $\mu_b$ is the pairon
chemical potential, $\delta_M$ the mini-gap, and noting the upper
cut-off energy $\Delta_0$.

\begin{figure}[b]
\centering
\includegraphics[trim=10 10 10 10, clip, width=6.0
cm]{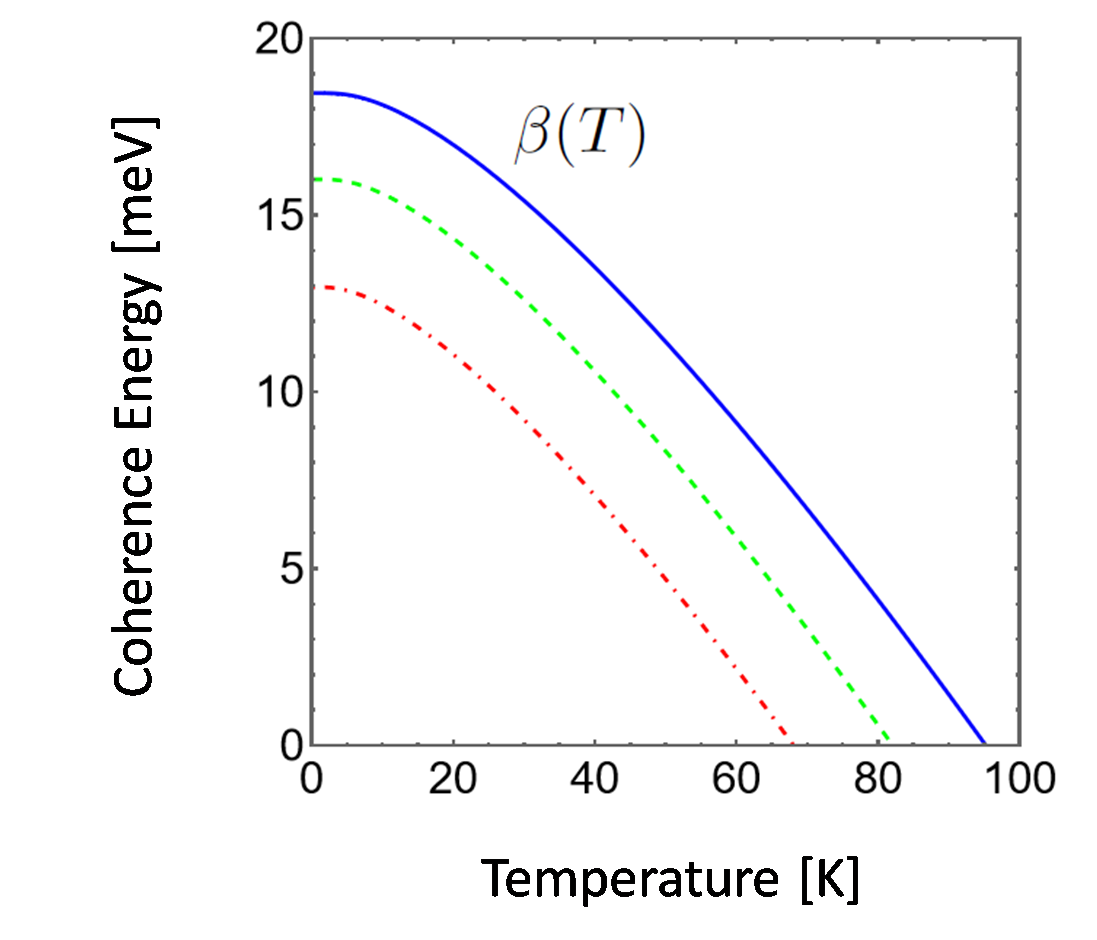} \caption{(Color online) Temperature dependence of
$\beta(T)$ using Eq.\,(\ref{Eq_bbeta}) for three different carrier
concentrations\,: optimally doped ($p=.16$, solid line), overdoped
($p=.2$, dashed lined), and underdoped ($p=.1$, dot-dashed line).}
\label{Fig bbeta}
\end{figure}
The temperature-dependent condensation energy is $\beta_c(T)=\beta_c\times n_{oc}(T)$, where $n_{oc}(T)$ is the normalized density of
condensed pairs and where $\beta_c$ is independent of temperature and only depends on the hole doping $p$. $\beta_c(T)$ can then be written:
\begin{equation}
    \beta_c(T) = \beta_c - \frac{\Gamma_{SC}}{\beta_c}\,
    \int_{\delta_M}^{\Delta_0}\,\overline{P_0}(\varepsilon_i)\, f_B(\varepsilon_i,T)\, d\varepsilon_i
\label{Eq_bbeta}
\end{equation}
where we take $\mu_b = 0$ consistent with $T\leq T_c$. We consider that
$\beta_c(T)$ is the correct order parameter of the SC transition
(see Fig.\,\ref{Fig bbeta}).

The integral in equation (\ref{Eq_bbeta}) can be evaluated
numerically to obtain $\beta_c(T)$ for any doping. However, an
analytic approach is satisfactory since $f_B(\varepsilon_i,T)$ is
very sharp at the lower integration value $\varepsilon_i \simeq
\delta_M$, wherein $\overline{P_0}(\varepsilon_i)$ can be taken out
of the integral. Furthermore, the critical temperature is defined by
the equation, $\beta_c(T_c)=0$ from which we obtain: $$\beta_c =
A\,\int_{\delta_M}^{\Delta_0}\, f_B(\varepsilon_i,T_c)\,
d\varepsilon_i$$ with $A = \frac{\Gamma_{SC}}{\beta_c}\,
\overline{P_0}(\delta_M)$. This leads to the analytical result, to a
good approximation\,:
\begin{equation}
\beta_c = \delta_M - A\,k_B\,T_c \ln(\exp(\delta_M/k_B T_c) - 1)
\label{Eq_beta_Tc}
\end{equation}
At optimal doping, using $\delta_M = 1.1\, p_{min}\, \beta_c=1.014$\,meV and $A=1.04$, we find the result for
the characteristic ratio \,: $$c=\frac{\beta_c}{k_B\,T_c} = 2.24$$ As shown in Fig.\,\ref{Fig_Tc}, as a function of $p$, the
calculated $T_c$ values using Eq.\,(\ref{Eq_beta_Tc}) match the
predicted line, $\propto p'(1-p')$, with very good accuracy (the
estimated uncertainty being $c=2.24\pm 0.03$). This confirms that
both the mini-gap $\delta_M$ and the correlation energy $\beta_c$
follow the $T_c$ dome from underdoped to overdoped sides of the dome.
% Figure 6

\begin{figure}[t]
\centering
\includegraphics[trim=10 10 10 10, clip, width=6.0
cm]{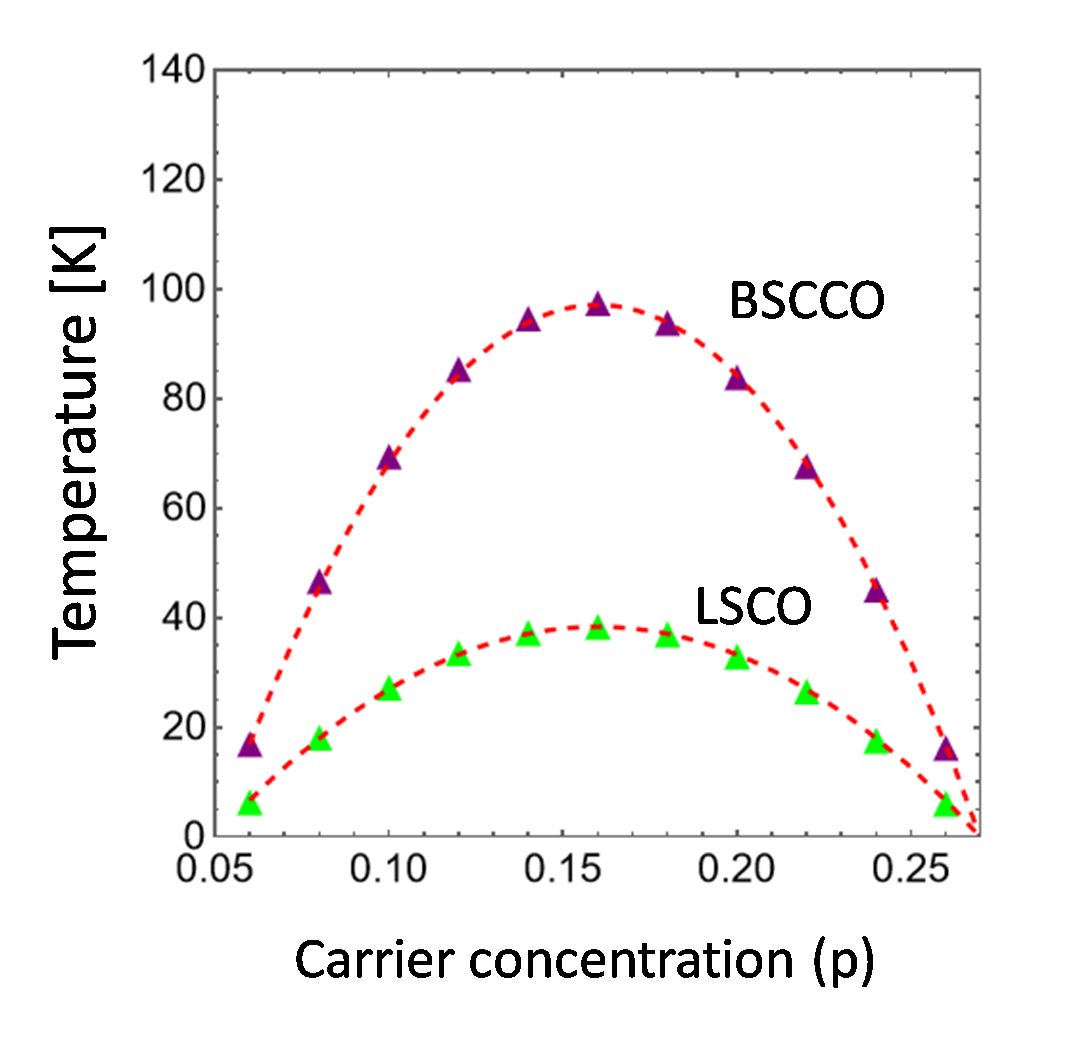} \caption{(Color online) Critical
temperature calculated using the theoretical equation
(\ref{Eq_beta_Tc}) for Bi$_2$Sr$_2$CaCu$_2$O$_{8+\delta}$ (purple
triangle) and La$_{2-x}$Sr$_x$CuO$_4$ (green triangles). Dashed
lines: $T_c(p)$ using the theoretical model
$T_c(p)=J_{eff}p'(1-p')/(2.24\,k_B)$, with $J_{eff} = 74$\,meV and
$J_{eff} = 31$\,meV in the upper and lower curves, respectively. In
the case of La$_{2-x}$Sr$_x$CuO$_4$, we neglect the `kink' in the
$T_c(p)$ curve on the underdoped side, due to a possible charge
ordering.} \label{Fig_Tc}
\end{figure}

\subsection{Discussion}

The pairon model expresses the critical temperature in terms of the
characteristic energy scale of the SC condensation, $\beta_c$. As
discussed previously, $\beta_c$, together with the other
characteristic energies, the energy gap $\Delta_p$ and the mini-gap
$\delta_M$, are proportional to the single energy scale $J_{eff}$.
Thus, both the pairing energy and the mutual
interaction between pairs arise from a single effect: the local
antiferromagnetic environment of the pairons.

It is important to note that the numerical factor
$\beta_c/k_BT_c\simeq 2.24$ has a very different physical origin
compared to the analogous BCS relation between the gap and the
critical temperature: $\Delta(0)/k_BT_c\simeq 1.76$. In our case it is
rather determined by the Bose-Einstein distribution, which governs
the pair excitations at finite temperature, and the presence of the
mini-gap in the excitation spectrum. As clearly
shown in the previous section, the cuprate $T_c$ equation is
proportional to both $J_{eff}$ and to $p'(1-p')$, the latter being due to
the square lattice constraint. These observations strongly support a
single mechanism governing the SC condensation all along the
$T_c$-dome.

The coupling of quasiparticles to virtual pair
excitations, explicit in the second term of the gap function,
$-\Gamma_{SC}\,P_0(E_k)$, not only renormalizes the Fermi level gap,
but provides a clear explanation of the unconventional QP-DOS shape
observed at low temperature. The signatures seen in the DOS, i.e.
the wide but sharp QP peaks followed by the dips, occur only in the
coherent state, with $\beta_c$ the underlying key parameter. This
points to a new all-electron mechanism leading to SC coherence.
Moreover, no collective mode nor retarding effects are involved in
the resulting pairon-pairon interaction.

It would be erroneous to conclude that {\it
thermal} pair-breaking quasiparticle excitations are absent in the
pairon model. Quasiparticle excitations, existing principally in the
range $T_c$ to $T^*$, are a key ingredient to describe the specific
heat \cite{SolstatCom_Noat2024}, the Fermi arcs near the nodal lines
\cite{EPL_Noat2019}, temperature-dependent tunneling
\cite{Jphys_Sacks2017} and ARPES \cite{Jphys_Sacks2018} experiments.
The pair-breaking quasiparticle effects, dependent on Fermi-Dirac
statistics, progressively increase as $T^*$ approaches $T_c$ on the
overdoped side of the dome. This does not contradict the present
condensation mechanism linked to the low-lying pairon excited
states, which follow Bose-Einstein statistics.

\subsection{Conclusions}

We have shown that the condensation mechanism in cuprates involves
the coupling of the SC condensate to the pair excited states leading
to an unconventional SC ground state. This coupling
results in an energy-dependent gap function, $\Delta(E)$, whose key
energy scale is the effective AF exchange energy, $J_{eff}$. This
provides strong evidence that the cuprate mechanism involves the AF
local environment of pairons. In contrast to spin-fluctuation or
phonon based pairing, $\Delta(E)$ has the characteristics of an
instantaneous (non-retarded) interaction, in agreement with recent
more microscopic models \cite{PRB_Zinni2021,NewJPhys_Yamase2023}.

The fundamental parameters of the cuprate ground
state\,: the condensation energy per pair, $\beta_c$, the energy gap
$\Delta_p$, and the width of the excited pair energy distribution,
$\sigma_0$, are proportional to the single energy scale, $J_{eff}$.
These key parameters are deduced from the quasiparticle spectrum
measured in tunneling experiments and agree quantitatively with the
predictions of the pairon model. Indeed, the strong coupling of
quasiparticles to virtual excited pairs explains the unconventional
`peak-dip-hump' shape  of the tunneling DOS. The SC ground state,
captured by the self-consistent gap function $\Delta(E)$ and the
shape of the quasiparticle spectrum, strongly suggest an
`all-electron' mechanism of SC phase coherence.

Turning to the {\it thermal} excited pairs, we obtained an
analytical form for the critical temperature. It revealed that
$\beta_c/(k_B\,T_c) \simeq 2.24$ is a universal constant throughout
the full hole-doping range. Moreover, we show that the temperature
dependance of $\beta_c(T)$ has the characteristics of an order
parameter of the transition, and not the pairing gap, $\Delta_p(T)$.
The latter is rather linked to the higher pseudogap temperature
$T^*\propto \Delta_p(0)$, consistent with the decay of excited
pairons above $T_c$, or when coherence is lost due to disorder or
within the vortex core\,\cite{EPL_Sacks2017, ModelSimul_Noat2023}.
Therefore, the BCS ratio $\Delta_p/(k_B\,T_c)$ has no particular
meaning in the case of cuprates.

Contrary to the BCS case, the numerical factor $\beta_c/(k_B\,T_c)$
is mainly dependent on the Bose-Einstein statistics governing pair
excitations and by the value of the mini-gap $\delta_M$ present in
the energy excitations. Since the result for $T_c$ is both
proportional to $J_{eff}$ and follows the square lattice constraint,
i.e. $p'(1-p')$, the same mechanism must govern the superconducting
condensation all along the $T_c$-dome.

The key role of the excited pairon states allows to better understand the $T$-dependence of the SC coherence energy
$\beta_c(T)$ and the transition to the pseudogap state above $T_c$.
Significant insight into the wide variety of major experiments is possible, such as\,: the temperature and magnetic phase diagrams, the tunneling spectra, the magnetic susceptibility, the specific
heat, and the nodal Fermi arcs. The exchange energy is shown to be
the key parameter, so that the criteria mentioned in the
introduction concerning a successful model, such as its basic
simplicity and the comparison to experiment, are potentially
fulfilled. Finally, the particular nature of the condensate-virtual
pair excitations described here can perhaps serve as a guide for
future more microscopic models.

\subsection{Acknowledgements}

The authors gratefully acknowledge discussions with Dr. Hiroshi
Eisaki, Dr. Shigeyuki Ishida (AIST, Tsukuba).

A.M. and W.S. acknowledge partial support from the French National
Research Agency (ANR), project `Superstrong' under contract no.
ANR-22-CE30-0010.

W.S. is grateful for continual support of the RIIS Institute of
Okayama University, Japan, to Prof. Takayoshi Yokoya (host), and his
`visiting professor' status while on leave from SU.


\begin{thebibliography}{99}
%\input{Biblio_topo_long_V01D}\Large\huge
%\input{Biblio_V0H}
%\bibitem{Corresp} $^*$ Corresponding author.\\
%E-mail address: yves.noat@insp.jussieu.fr (Y. Noat).}

\bibitem{PhysicaC_Singh2021} Navinder Singh, Leading theories of the
cuprate superconductivity: A critique, Physica C {\bf  580}, 1353782
(2021).

\bibitem{Occam} Occam Summa logicae (Sum of Logic).

\bibitem{Popper1959}
K.~R.~Popper, \textit{The Logic of Scientific Discovery},
Hutchinson, London (1959). [Originally published in german as
\textit{Logik der Forschung}, 1934.] Note: K.P. chose the wording
`falsifiable' in lieu of `refutable'.

\bibitem{Note_Popper} footnote: KP used the unfortunate wording
`falsifiable'.

\bibitem{PR_BCS1957} J. Bardeen, L. Cooper, J.R. Schrieffer, Theory of Superconductivity, Phys. Rev. {\bf 108} 1175 (1957).

\bibitem{Parks} R.~D.~Parks (ed.), \textit{Superconductivity}, Vols.~1 and 2, Marcel
Dekker, New York (1969).


\bibitem{PR_Cooper1956} Leon N. Cooper, Bound electron pairs in a degenerate Fermi gas, Physical Review {\bf 104}, 1189-1190 (1956).

\bibitem{PR_Giaever1962} I. Giaever, H. R. Hart, Jr., and K. Megerle, Tunneling into superconductors at temperatures below 1$\,^{\circ}\mathrm{K}$, Phy. Rev. {\bf 126}, 941 (1962).

\bibitem{PR_Douglass} D. H. Douglass and R. Meservey, Energy gap measurements by tunneling between superconducting films. I. temperature dependence, Phys. Rev. A {\bf  135}, 19 (1964).

\bibitem{PR_Brown1953} A. Brown, M. W.Zemansky, and H. A. Boorse, The superconducting and normal heat capacities of niobium, Phys. Rev. {\bf 92}, 52 (1953).

\bibitem{PR_Corak1956} W. S. Corak, B. B. Goodman, C. B. Satterthwaite, and A. Wexler, Atomic heats of normal and superconducting vanadium, Phys. Rev. {\bf 102}, 656 (1956)

\bibitem{PR_Maxell1950} Emanuel Maxwell, Isotope effect in the superconductivity of mercury, Phys. Rev. {\bf 78}, 477 (1950).

\bibitem{PR_Reynolds1950} C. A. Reynolds, B. Serin, W. H. Wright, and L. B. Nesbitt, Superconductivity of isotopes of mercury, Phys. Rev. {\bf 78}, 487 (1950).

\bibitem{PRL_McMillan1965} W. L. McMillan and J. M. Rowell, Lead phonon spectrum calculated from superconducting density of states, Phys. Rev. Lett. {\bf 14}, 108 (1965).

\bibitem{ZPhys_Bednorz1986} J. G. Bednorz, K. A. M\"{u}ller, Possible high $T_c$ superconductivity in the Ba--La--Cu--O system , Zeitschrift f\"{u}r Physik B Condensed Matter {\bf 64}, 189 (1986).

\bibitem{Sci_Anderson1987} P. W. Anderson, The Resonating valence bond
state in La$_2$CuO$_4$ and superconductivity, Science {\bf  235},
1196 (1987).

\bibitem{Nat_Emery}V. Emery et S. Kivelson, Importance of phase fluctuations in superconductors with small superfluid density, Nature  {\bf 374}, 434 (1995)

\bibitem{PRB_varma}C. Varma, Non-Fermi-liquid states and pairing instability of a general model of copper oxide metals, Phys. Rev. B {\bf  55}, 14554 (1997).

\bibitem{PRL_Huscroft1998} Carey Huscroft and Richard T. Scalettar,
Evolution of the density of states gap in a disordered
superconductor, Phys. Rev. Lett. {\bf 81}, 2775 (1998).

\bibitem{PRL_Curty2002} Philippe Curty and Hans Beck, Thermodynamics
and phase diagram of high temperature superconductors, Phys. Rev.
Lett. {\bf 91}, 257002 (2003).

\bibitem{JPhysConf_Newns2007} D. M. Newns and C. C. Tsuei, Fluctuating
bond model of high temperature superconductivity in cuprates, J.
Phys.: Conf. Ser. {\bf 92}, 012007 (2007).

\bibitem{JChemPhys_Tahir-Kheli2010} Jamil Tahir-Kheli and William A.
Goddard, III, Universal properties of cuprate superconductors: $T_c$
Phase diagram, room-temperature thermopower, neutron spin resonance,
and STM incommensurability explained in terms of chiral plaquette
Pairing, J. Phys. Chem. Lett.  {\bf 1}, 1290 (2010).

\bibitem{PRL_Gull2013} E. Gull, O. Parcollet and A. J. Millis,
Superconductivity and the pseudogap in the two-dimensional Hubbard
model, Phys. Rev. Lett. {\bf 110}, 216406 (2013).

\bibitem{RevMod_Chen2024}Qijin Chen, Zhiqiang Wang, Rufus Boyack,
Shuolong Yang, and K. Levin, When superconductivity crosses over:
From BCS to BEC, Rev. Mod. Phys. {\bf 96}, 025002 (2024).

\bibitem{PhysicaB_Marino2025} E.C. Marino, The phase diagram of
High-$T_c$ cuprates, Physica B: Condensed Matter {\bf 699}, 416815
(2025).

\bibitem{PRL_Fang2006}  A. C. Fang, L. Capriotti, D. J. Scalapino, S. A. Kivelson, N. Kaneko, M. Greven, and A. Kapitulnik, Gap-inhomogeneity-induced electronic states in superconducting, Bi$_2$Sr$_2$CaCu$_2$O$_{8+\delta}$, Phys. Rev. Lett. {\bf 96}, 017007(2006).

\bibitem{PRL_Cren2000}
T. Cren, D. Roditchev, W. Sacks, and J. Klein, J.-B. Moussy, C.
Deville-Cavellin, and M. Lagu\"es, Influence of disorder on the
local density of states in high-$T_c$ superconducting thin films,
 Phys. Rev. Lett.  {\bf 84}, 147(2000).

 \bibitem{Revmod_Fisher2007} \O. Fischer, M. Kugler, I. Maggio-Aprile,
C. Berthod and C. Renner, Scanning tunneling spectroscopy of the
cuprates, Rev. Mod. Phys. {\bf 79}, 353 (2007).

\bibitem{PRL_Ahmadi2011} O. Ahmadi, L. Coffey, and J. F. Zasadzinski, N. Miyakawa, L. Ozyuzer, Eliashberg Analysis of Tunneling experiments: support for the pairing glue hypothesis in cuprate superconductors, Phys. Rev. Lett. {\bf 106}, 167005 (2011).

\bibitem{PRB_Berthod2013} C. Berthod, Y. Fasano, I. Maggio-Aprile, A. Piriou, E. Giannini, G. Levy de Castro, and  \O. Fischer, Strong-coupling analysis of scanning tunneling spectra in Bi$_2$Sr$_2$CaCu$_2$O$_{10+\delta}$, Phys. Rev. B {\bf 88}, 014528 (2013).

\bibitem{PRL_Pan2000} S. H. Pan, E. W. Hudson, A. K. Gupta, K.-W. Ng, H. Eisaki, S. Uchida, and J. C. Davis, STM Studies of the electronic structure of vortex cores in Bi$_2$Sr$_2$CaCu$_2$O$_{8+\delta}$, Phys. Rev. Lett. {\bf 85}, 1536 (2000).

\bibitem{PRL_renner1998_T} Ch. Renner,B. Revaz, J.-Y. Genoud, K. Kadowaki,and {{\O}}. Fischer, Pseudogap precursor of the superconducting gap in under- and overdoped Bi$_2$Sr$_2$CaCu$_2$O$_{8+\delta}$, Phys. Rev. Lett., {\bf 80} 149 (1998).

\bibitem{JphysSocJap_Sekine2016} R. Sekine, S. J. Denholme, A. Tsukada, S. Kawashima, M. Minematsu,T. Inose, S. Mikusu, K. Tokiwa, T. Watanabe, and N. Miyakawa, Characteristic features of the mode energy estimated from tunneling conductance on TlBa$_2$Ca$_2$Cu$_3$O$_{8.5+\delta}$, J. Phys. Soc. Jpn. {\bf 85}, 024702 (2016).

\bibitem{JPhysSocJap_Nakano1998} Tohru Nakano, Naoki Momono, Migaku Oda, and Masayuki Ido, Correlation between the Doping Dependences of Superconducting Gap Magnitude $2\Delta_0$ and Pseudogap Temperature $T^*$ in High-T$_c$ Cuprates , J. Phys. Soc. Jpn. {\bf 67}, 2622-2625 (1998).

\bibitem{RepProgPhys_Hufner2008} S. H\"ufner, M. A. Hossain, A Damascelli, and G. A. Sawatzky,Two gaps make a high-temperature superconductor?,
Rep. Prog. Phys., {\bf 71}, 062501 (2008).

\bibitem{PRB_Sacks2006} W. Sacks, T. Cren, D. Roditchev, and B. Dou\c{c}ot, Quasiparticle spectrum of the cuprate Bi$_2$Sr$_2$CaCu$_2$O$_{8+\delta}$: Possible connection to the phase diagram, Phys. Rev. B  {\bf 74}, 174517(2006).

%%%%%%%%%%%%%%% 35

\bibitem{EPL_Wen2003} H. H. Wen, H. P. Yang, S. L. Li, X. H. Zeng, A. A. Soukiassian, W. D. Si and X. X. Xi, Hole doping dependence of the coherence length in
La$_{2-x}$Sr$_x$CuO$_4$ thin films,  Europhys. Lett., {\bf  64 },
790 (2003).

\bibitem{EPL_Wang2008} Y. Wang and H.-H. Wen, Doping dependence of the upper critical field
in La$_{2-x}$Sr$_x$CuO$_4$ from specific heat, Europhys. Lett., {\bf
81}, 57007 (2008).

%%%%%%%%%%% 38

\bibitem{PhysicaC_Uemura1997} Y. J. Uemura, Bose-Einstein to BCS crossover picture for high-$T_c$ cuprates, Physica C {\bf 282-287}, 194-197 (1997).

\bibitem{PRL_Miyakawa1999}  N. Miyakawa, J. F. Zasadzinski, L. Ozyuzer, P. Guptasarma, D. G. Hinks, C. Kendziora, and K. E. Gray, Predominantly superconducting origin of large energy gaps in underdoped Bi$_2$Sr$_2$CaCu$_2$O$_{8+\delta}$ from tunneling spectroscopy, Phys. Rev. Lett.  {\bf 83}, 1018 (1999).

\bibitem{SciTech_Sacks2015}  W. Sacks, A. Mauger, Y. Noat, Pair\,--\,pair interactions as a mechanism for high-T$_c$ superconductivity, Superconduct. Sci. Technol., {\bf 28}
105014 (2015).

\bibitem{PRB_Zinni2021} Luciano Zinni, Matias Bejas, and Andr\'es Greco, Superconductivity with and without glue and the role of the double-occupancy forbidding constraint in the $t-J-V$ model, Phys. Rev. B {\bf 103}, 134504 (2021).

\bibitem{NewJPhys_Yamase2023} Hiroyuki Yamase, Spin-fluctuation glue disfavors high-critical temperature of superconductivity?,
New J. Phys. {\bf 25}, 083049 (2023).

\bibitem{RevMod_Dagatto1994} Elbio Dagotto, Correlated electrons in high-temperature superconductors, Rev. Mod. Phys. {\bf 66}, 763 (1994).

\bibitem{EPL_Sacks2017} W. Sacks, A. Mauger and Y. Noat, Cooper pairs without glue in high-$T_c$ superconductors: A universal phase diagram, Euro. Phys. Lett {\bf 119}, 17001 (2017).

\bibitem{PRB_Cyr-Choiniere2018} O. Cyr-Choini\`re, R. Daou, F. Lalibert\'e, C. Collignon, S. Badoux, D. LeBoeuf, Pseudogap temperature $T^*$ of cuprate superconductors from the Nernst effect ,  Phys. Rev. B {\bf 97}, 064502 (2018).

\bibitem{PhysLettA_Noat2022} Yves Noat, Alain Mauger, William Sacks. Superconductivity in cuprates governed by topological constraints. Physics Letters A {\bf 444}, 128227 (2022).

\bibitem{ModelSimul_Noat2023} Yves Noat  Alain Mauger, William Sacks, Statistics of the cuprate pairon states on a square lattice, Modelling Simul. Mater. Sci. Eng. {\bf 31}, 075010 (2023).

\bibitem{Jphys_Sacks2017} William Sacks, Alain Mauger and Yves Noat, Universal spectral signatures in pnictides and cuprates: the role of quasiparticle-pair coupling, J. Phys.: Condens. Matter  {\bf 29}, 445601 (2017).

\bibitem{EPJB_Sacks2016} William Sacks, Alain Mauger, and Yves Noat, Unconventional temperature dependence of the cuprate excitation spectrum, Eur. Phys. J. B {\bf 89}, 183 (2016).

\bibitem{Jphys_Sacks2018} William Sacks, A. Mauger and Y. Noat, Origin of the Fermi arcs in cuprates: a dual role of quasiparticle and pair excitations, Journal of Physics: Condensed Matter, {\bf 30},  475703 (2018).

\bibitem{EPL_Noat2019} Yves Noat, Alain Mauger and William Sacks, Single origin of the nodal and antinodal gaps in cuprates, Euro. Phys. Lett {\bf 126}, 67001 (2019).

\bibitem{SolStatCom_Noat2021} Y. Noat, A. Mauger, M. Nohara, H. Eisaki, W. Sacks
, How `pairons' are revealed in the electronic specific heat of cuprates, Solid State Communications {\bf 323}, 114109 (2021).

 \bibitem{SolstatCom_Noat2024}Yves Noat, Alain Mauger, William Sacks, Unraveling pairon excitations and the antiferromagnetic contributions in the cuprate specific heat, Solid State Communications {\bf 394}, 115707 ( 2024).

\bibitem{PhysLettA_Noat2025} Yves Noat, Alain Mauger, William Sacks.
Magnetic phase diagram of cuprates and universal scaling laws, Physics Letters A {\bf 544}, 130460 (2025).

\bibitem{Economou} This expression for the excited states, wherein the energy lies within an energy band of finite width, is adapted from a discussion in the book: E. N. Economou, Green's Functions in Quantum Physics, Springer (2006).

\bibitem{EPL_Cren2000} T. Cren, D. Roditchev, W. Sacks and J. Klein, Constraints on the quasiparticle density of states in high-$T_c$ superconductors, Europhys. Lett. {\bf 52}, 203 (2000).

\bibitem{PhysicaC_Hoogenboom2000}  B.W. Hoogenboom, Ch. Renner, B. Revaz, I. Maggio-Aprile, \O. Fischer  Low-energy structures in vortex core tunneling spectra in
 Bi$_2$Sr$_2$CaCu$_2$O$_{8+\delta}$, Physica C {\bf 332} 440 (2000).


%\bibitem{SciTech_Sacks2015}  W. Sacks, A. Mauger, Y. Noat, Pair\,--\,pair interactions as a mechanism for high-T$_c$ superconductivity, Superconduct. Sci. Technol., {\bf 28}, 105014 (2015).

\bibitem{PRL_Renner1998} Pseudogap precursor of the superconducting gap in under- and overdoped Bi$_2$Sr$_2$CaCu$_2$O$_{8+\delta}$, Ch. Renner, B. Revaz, J.-Y. Genoud, K. Kadowaki, and \O. Fischer, Phys. Rev. Lett. {\bf 80}, 149 (1998).

\bibitem{Sci_McElroy2005} K. McElroy, J. Lee, J.A. Slezak, D.-H. Lee, H. Eisaki, S. Uchida, J.C. Davis, Atomic-scale sources and mechanism of nanoscale electronic disorder in Bi$_2$Sr$_2$CaCu$_2$O$_{8+\delta}$, Science {\bf 309}, 1048 (2005).

\bibitem{PRL_Zasadzinski2001}
J. F. Zasadzinski, L. Ozyuzer, N. Miyakawa, K. E. Gray, D. G. Hinks, and C. Kendziora,
Correlation of tunneling spectra in Bi$_2$Sr$_2$CaCu$_2$O$_{8+\delta}$ with the resonance spin excitation, Phys. Rev. Lett. {\bf 87}, 067005 (2001).

\bibitem{Science_YHe}Fermi surface and pseudogap evolution in a cuprate
superconductor, Yang He, Yi Yin, M. Zech, Anjan Soumyanarayanan,
Michael M. Yee, Tess Williams, M. C. Boyer, Kamalesh Chatterjee, W.
D. Wise, I. Zeljkovic, Takeshi Kondo, T. Takeuchi, H. Ikuta, Peter
Mistark, Robert S. Markiewicz, Arun Bansil, Subir Sachdev, E. W.
Hudson, and J. E. Hoffman, Science, {\bf 344}, 608 (2014).

\bibitem{LowtempPhys_Ido1999} M. Ido, N. Momono and M. Oda, Correlation between Superconducting gap and pseudogap in high-$T_c$ cuprates, J. Low Temp. Phys. 117, {\bf 329} (1999).

\bibitem{JPhysSocJap_Dipasupil2002} R. M. Dipasupil, M. Oda, N. Momono, and M. Ido, The Physical Society of Japan Energy Gap Evolution in the Tunneling Spectra of Bi$_2$Sr$_2$CaCu$_2$O$_{8+\delta}$, Journal of the Physical Society of Japan {\bf 71}, 1535 (2002).

%\bibitem{PhysicaC_Hoffman2003} J.E\^{A} Hoffman, E.W Hudson, K.M\^{A} Lang, H\^{A} Eisaki, S\^{A} Uchida, J.C\^{A} Davis,
%Vortex-induced quasi-particle `checkerboard' in Bi$_2$Sr$_2$CaCu$_2$O$_{8+\delta}$, Physica C {\bf 388-389}, 703 (2003).

\end{thebibliography}
\end{document}